# Coupling Transitions in Helical Polymers: The Case of the Helix-Coil and Coil-Globule Transitions

*Vikas Varshney and Gustavo A. Carri*[*]

The Maurice Morton Institute of Polymer Science, The University of Akron, Akron, OH 44325-3909

ABSTRACT

We explore the coupling between the helix-coil and coil-globule transitions of a helical polymer using Monte Carlo simulations. A very rich state diagram is found. Each state is characterized by a specific configuration of the chain which could be a helix, a random coil, an amorphous globule or one of various other globular states which carry residual helical strands. We study the boundaries between states and provide further insight into the physics of the system with a detailed analysis of the order parameter and other properties.

---

[*] To whom any correspondence should be addressed. Electronic mail: gac@uakron.edu

Proteins exist in one of three possible states: as a collapsed globule, as a random coil or in a well-defined native state.[1] Transitions among these configurations can be induced by the application of an external stimulus like a decrease in temperature or solvent quality. Among all the possible transitions, one stands out due to its common occurrence in biological and synthetic macromolecules: the coil-globule transition.[2] During this transition, the macromolecule changes its configuration from a random coil to a compact globule upon a decrease in temperature. However, while this transition is continuous for synthetic polymers, the case is more complex for biopolymers like proteins[3] where residual secondary structures persist even in the globular state.[4] This residual ordering is a consequence of a competition between the native folded state and the collapsed amorphous globular structure, both being possible equilibrium states at low temperatures. Hence, understanding this competition between various possible equilibrium, low-temperature states in proteins is a first class challenge in polymer physics. However, due to the structural complexity of proteins, we restrict ourselves to the study of simpler molecular structures. We focus on the case of homopolypeptides which are known to form α-helices and undergo the helix-coil transition.[5,6] Consequently, the main purpose of this letter is to study the coupling of the helix-coil and coil-globule transitions in homopolypeptides, problem that belongs to the more general classification of possible couplings of transitions in macromolecular systems proposed by Di Marzio.[7] This problem is also similar to the collapse transition of polymers with variable stiffness studied by Stukan *etal.* recently.[8] However, it should be noted that the latter paper adopts a purely bending potential for introducing chain stiffness, which is not identical to the torsional potential adopted in the present letter which we describe below.

The polypeptide is described with a coarse-grained model which maps the polymer onto a freely rotating chain where each residue is represented by a bead.[6] The cooperative nature of the helix-coil transition is captured using a geometric criterion based on the concept of "torsion" of a curve. This cooperativity is known to emerge from the formation of hydrogen bonds between pairs of residues $i$ and $i+4$ which, in turn, constrain the spatial positions and orientations of residues $i+1$, $i+2$ and $i+3$. Explicitly, the torsion determines the conformation (helix or coil) of each bead. If the torsion of a bead differs from the one of the helix by less than a cutoff value then the bead is in the helical state otherwise, it is part of a random coil which is the reference state. Helical beads carry a negative enthalpy, $C$ (=-1300K),[6] that models the formation of a hydrogen bond and determines the transition temperature of the helix-coil transition, $T_{hc}$; 312K for our case. These concepts were implemented in a Monte Carlo simulation based on the Wang-Landau (WL) algorithm and their consequences were explored.[6,9]

In this letter, we generalize the model to account for the coil-globule transition in a similar manner to previous works[8,10] and explore the physical consequences of the simultaneous presence of both transitions. This is accomplished through an implicit-solvent Monte Carlo simulation based on the WL algorithm which computes the density of states (DOS) of the system by sampling all the phase space. Solvent effects are modeled as an effective interaction between beads described by a modified LJ potential of the form

$$V_{ij} = 4\varepsilon\left(\left(\frac{\sigma}{r_{ij}}\right)^{12} - \left(\frac{\sigma}{r_{ij}}\right)^{6}\right) \qquad r_{ij} \geq \sigma, \qquad (2b)$$

$$V_{ij} = \infty \qquad r_{ij} < \sigma, \qquad (2b)$$

where $r_{ij}$ is the distance between beads $i$ and $j$, $\varepsilon$ is the strength of the interaction and $\sigma$ the bead diameter. The effect of solvent is taken into account via a *compactness* parameter $\lambda$ defined as $\lambda = -\sum_{i}\sum_{j>i} V_{ij}/\varepsilon$. Hence, a more compact configuration corresponds to a larger value of $\lambda$. For each configuration of the chain we record the number of beads in the helical state, $N_s$, and the value of $\lambda$. Therefore, the energy of the configuration is

$$E = C.N_s - \varepsilon.\lambda \quad . \tag{3}$$

Since $E$ is a function of $N_s$ and $\lambda$, we define the DOS of the system as a function of same parameters, i.e., $g(N_s,\lambda)$ and use the WL algorithm to compute it. Once $g(N_s,\lambda)$ is known, the equilibrium properties are analyzed using formulas of statistical mechanics.

Figure 1 shows the state diagram for a polymer with 60 beads together with typical configurations of the chain. The dark and light segments represent helical and coil beads, respectively. The boundaries between states were obtained from the peak in the heat capacity. Figure 1 shows some boundaries (small peaks) ending before merging with the other boundaries (large peaks). This occurs when the smaller peaks approach the large ones; they become shoulders or are absorbed by the large peaks hence, they could not be resolved. At low temperatures and for small $\varepsilon$ the chain is in the helical state which, upon an increase in temperature, undergoes the helix-coil transition.[5,6] Moreover, at low temperatures an increase in $\varepsilon$ results in a cascade of *continuous conformational* transitions starting in the helical state and ending in a collapsed globular state with residual local order as depicted by the dark helical strands. In addition, for any $\varepsilon$, the residual order is melted by an increase in temperature. Indeed, at high temperatures, the chain adopts either a random coil or an amorphous globular state, which are separated by the coil-globule transition boundary. Figure 1 also shows how the transition temperature changes with $\varepsilon$.

Figures 2, 3 and 4 show the mean square radius of gyration, $\langle R_g^2 \rangle$, order parameter of the helix-coil transition, i.e. the fraction of helical beads, $\langle \theta \rangle$, and the number of helical strands,

$\langle N_H \rangle$, for values of temperature, T, and ε in the intervals (50K, 500K) and (0,1000), respectively. We have explored this large region of parameter space to identify as many states as possible.

When ε=0, the helix-coil transition is recovered.[5,6] Figure 2 shows the decrease in $\langle R_g^2 \rangle$ with increasing temperature suggesting a cooperative transition. This is confirmed by Fig. 3 which shows the typical sigmoid-type shape of $\langle \theta \rangle$ as a function of temperature, implying its cooperative nature. The transition occurs at $T_{hc}$ which is mostly determined by the parameter $C$.[6] At low temperatures, the helical content is one implying an all-helical configuration but, as the temperature increases the helical content decreases to zero indicating the melting of the helical strands. Figure 4 shows $\langle N_H \rangle$ as a function of temperature. The peak around $T_{hc}$ indicates that the helical structure breaks into more than one helical strand during the transition, in agreement with helix-coil transition theories.[5]

Figure 2 also shows that upon a small increase in ε, $\langle R_g^2 \rangle$ decreases. This decrease is clearly observed at high temperature and is a consequence of the stronger attraction among beads which tries to collapse the chain irrespective of temperature. At low temperatures the decrease in $\langle R_g^2 \rangle$ is not observed because the chain prefers the helical conformation.

At high temperatures the chain does not form helices and the model predicts a transition from the random coil to the globular state as ε increases (Fig. 2). This transition should also occur with decreasing temperature at fixed ε. Figure 2 shows an initial decrease of $\langle R_g^2 \rangle$ with decreasing temperature for temperatures higher than $T_{hc}$ and ε~200. However, the formation of the helical structure interferes with the globular state at temperatures close to $T_{hc}$.

At low temperatures the model predicts a cascade of step-like, *continuous conformational* transitions with increasing ε. The sharpness of these transitions is a result of the competition between various stable and meta-stable states present at low temperatures. Moreover, it indicates that each transition is cooperative as the helix-coil transition. This observation agrees with the experiments of Moore and collaborators who studied the helix-coil transition of phenylene ethynylene oligomers at constant temperature as a function of solvent quality and observed the typical sigmoid-type shape of ⟨θ⟩ which indicates cooperativity.[11] However, they observed only one transition which is due to the low degree of polymerization (DP) used in their studies. We also studied short chains (10 beads) and observed only one transition. However, longer chains, e.g. DP=20, 30, 45 and 60, showed a cascade of step-like transitions as discussed below.

For small values of ε and temperature, the chain is in the helical state (Fig. 3). However, as ε increases, ⟨θ⟩ decreases in a step-like manner at very specific values of ε. This decrease is due to the breaking of the helical structure caused by the LJ attractions and was observed for all the chain lengths studied (20, 30, 45 and 60 beads). This mechanism is similar to the one present in the standard helix-coil transition of long enough chains where the polymer undergoes the transition by breaking the all-helical conformation into shorter helical strands as the temperature is increased. However, the helix-coil transition does not consider solvent effects and the helical strands do not aggregate like in our case. Since the underlying mechanisms for breaking the helical structure are the same, cooperativity is expected to occur in the various, low-temperature transitions. Along the same lines, Fig. 2 shows a step-like decrease in $\langle R_g^2 \rangle$ for the same values of ε implying that the breaking of the helix also changes the dimensions of the chain, making it more compact. In addition, Fig. 4 shows that ⟨$N_H$⟩ also changes at each transition. First it increases from one to six in steps and, afterward, decreases to five at higher values of ε. These changes in $\langle N_H \rangle$ along with the decrease in $\langle \theta \rangle$ and $\langle R_g^2 \rangle$ support the proposed mechanism for breaking the helical structure illustrated in Fig. 1 with several snapshots of the polymer chain.

During the first step-like transition, the all-helical conformation breaks into two smaller, but relatively long, helical strands which may have any relative orientation in space. However, we argue that the most probable configuration is the one where they are aligned parallel to each other and are located side by side, as shown in Fig. 1. This relative alignment is favored by the increase in the number of contacts between the beads of both helical strands which decreases the LJ energy and the helical conformation of both strands which decreases the torsional energy. This first transition was observed for all the DP studied making it a characteristic feature of the model. An increase of ε generates a transition to a conformational state where three helical strands are parallel to each other which was also observed for all the chain lengths studied. As ε increases further, the model predicts states with 4 and 6 helical strands parallel to each other. However, the parallel alignment of the helices is disturbed by the large number of coil beads at higher values of ε. For example, for a 60-bead chain large values of ε generate a transition between two states both with six helical strands. In this case, the latter conformation with lower $\langle \theta \rangle$ has a lower degree of alignment of the strands, supporting the previous argument. A further increase of ε leads to a decrease to 5 helical strands in a non-parallel configuration. Similar behavior was observed for 45 beads. Interestingly, for the 30-bead case the residual helical strands were found to be parallel to each other. The same energetic argument used to rationalize the formation of the first state is also applicable to the other states until the LJ attractive interactions overwhelm the formation of helices and control the behavior of the chain.

We also studied the effect of chain length on the state diagram and found significant changes only in the low temperature regime. An increase in chain length has two main consequences. First, new states with more helical strands appear between the all-helical conformation and the collapsed globular state with residual order. Therefore, more and sharper transitions are observed. However, due to finite size effects, they are not transitions in the *thermodynamic* sense. This dependence on chain length is opposite to the one reported by Stukan and collaborators for semiflexible chains where an increase in chain length reduces the number of stable states of the chain (See Fig. 1 in Ref. 8) because no disk-like globular structures are present, only toroidal globules are predicted. These toroidal states are akin to the various globular states reported in this letter. Indeed, several toroidal states with different winding numbers can be found with increasing chain length and strength of the attractive interactions.[12] Second, the LJ attractive interactions become stronger (they increase as $DP^2$) and overwhelm the enthalpy gain due to the formation of longer or more helices (it increases as DP). Thus, smaller values of $\varepsilon$ are required to break the various structures. For example, for a 30-bead chain at 200K the all-helical conformation breaks when $\varepsilon \approx 200$ but, for a 60-bead chain the same transition occurs when $\varepsilon \approx 100$.

At low temperatures $\langle N_H \rangle$ increases with increasing $\varepsilon$ due to the breaking of the helices into shorter helical strands and then decreases due to the monotonic decrease in $\langle \theta \rangle$ (Figs. 3 and 4). This leads to a maximum in $\langle N_H \rangle$ which is determined by the chain length. In our case, this happens when $\langle N_H \rangle$ is equal to 3, 5 and 6 for chain lengths of 30, 45 and 60. However, $\langle N_H \rangle$ does not approach zero on further increase of $\varepsilon$ implying that some residual secondary structure remains even in the collapsed globular state. This residual structure melts with increasing temperature (Fig. 1). However, transitions from one globular state to another one with different number of helical strands are possible (Fig. 1) because the transition temperatures depend on $\varepsilon$. The lower the transition temperature, the more stable the helical structure. Hence, the stronger the interaction needed to break the helical strands which results in a transition at larger values of $\varepsilon$.

In summary, we have investigated the coupling between the helix-coil and coil-globule transitions. We found that, in addition to the helix-coil and coil-globule transitions, the model predicts a series of helix-globule and globule-globule transitions at low temperatures where an all-helical state collapses into a globular state with residual secondary structures through a cascade of continuous conformational transitions. Furthermore, the residual order disappears gradually with an increase in temperature leading to transitions between collapsed globular states and the *amorphous* globular or random coil states.

This material is based upon work supported by the National Science Foundation, Grant CHE-0132278. We also acknowledge The Ohio Board of Regents, Action Fund, Grant# R566.

LIST OF FIGURES

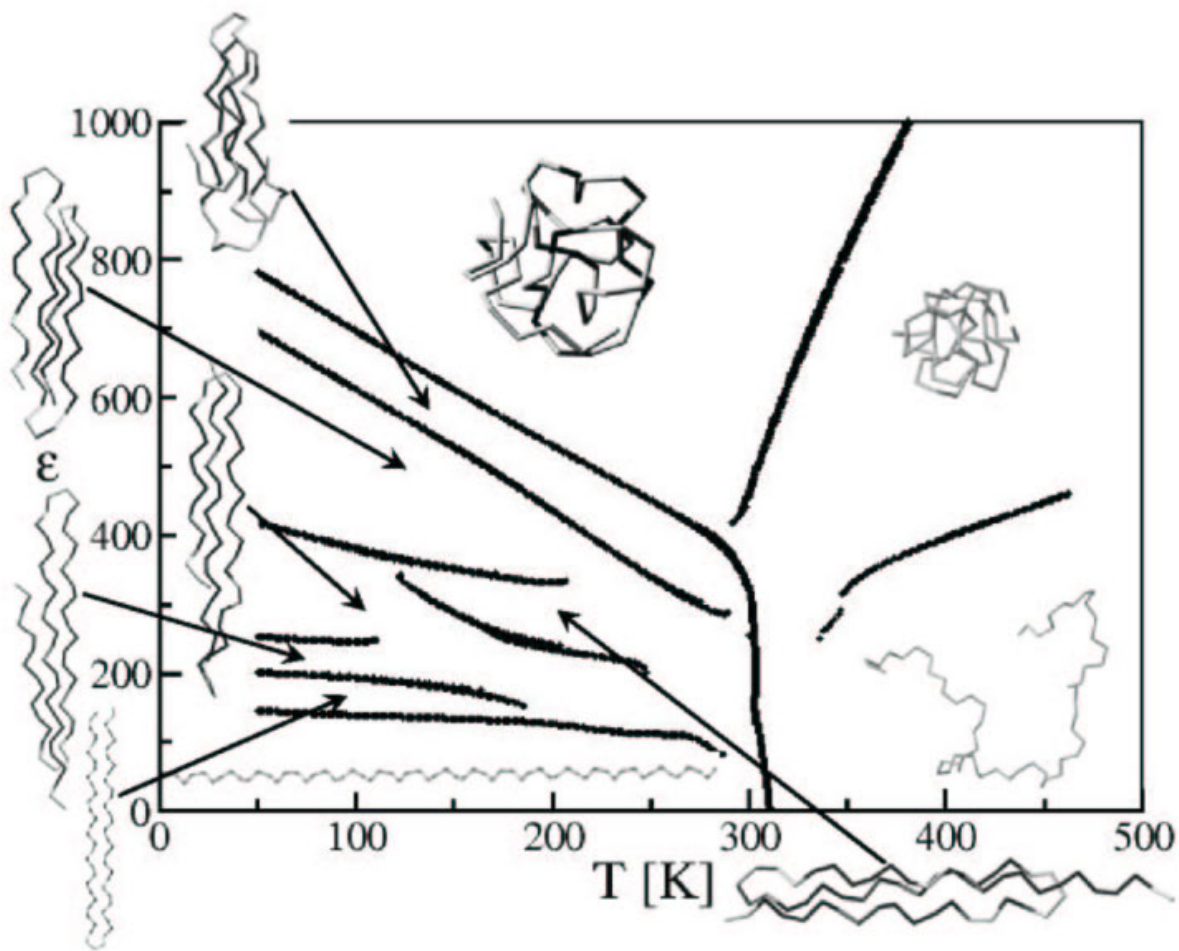

**Figure 1.** State diagram for a chain with 60 beads. Helical beads (dark segments), coil beads (light segments).

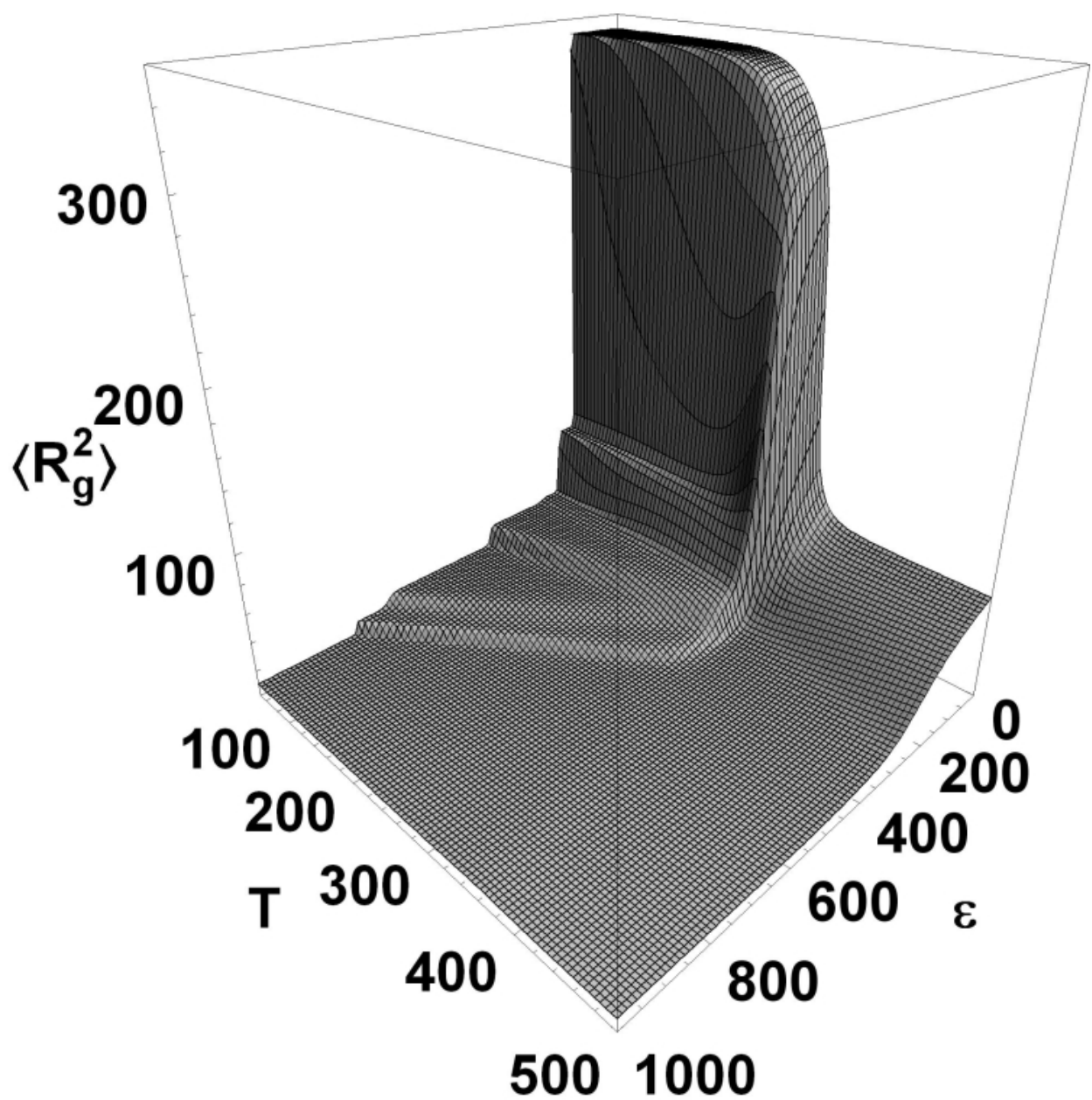

**Figure 2.** $\langle R_g^2 \rangle$ as a function of ε and temperature for a chain with 60 beads.

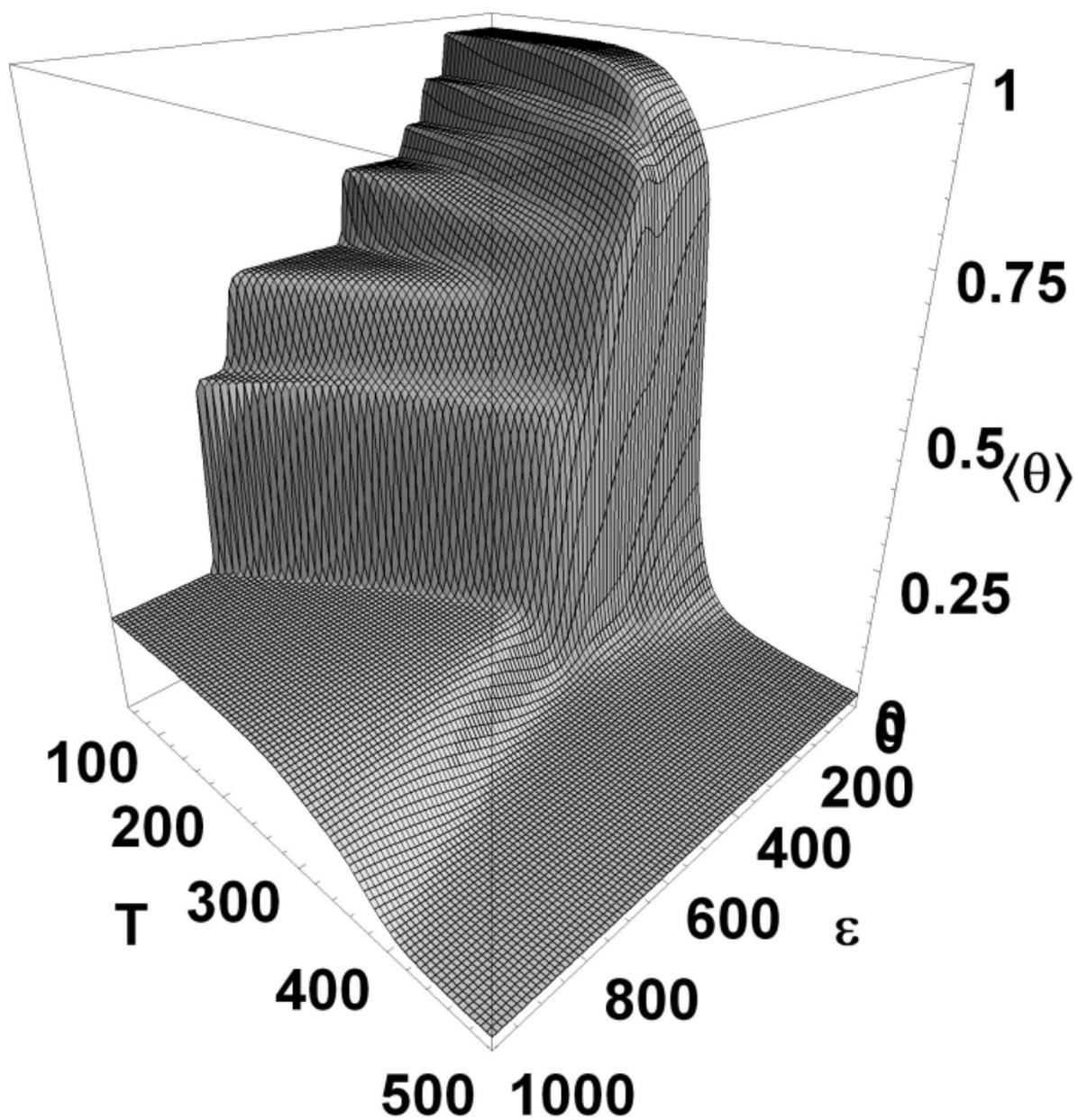

**Figure 3.** ⟨θ⟩ as a function of ε and temperature for a chain with 60 beads.

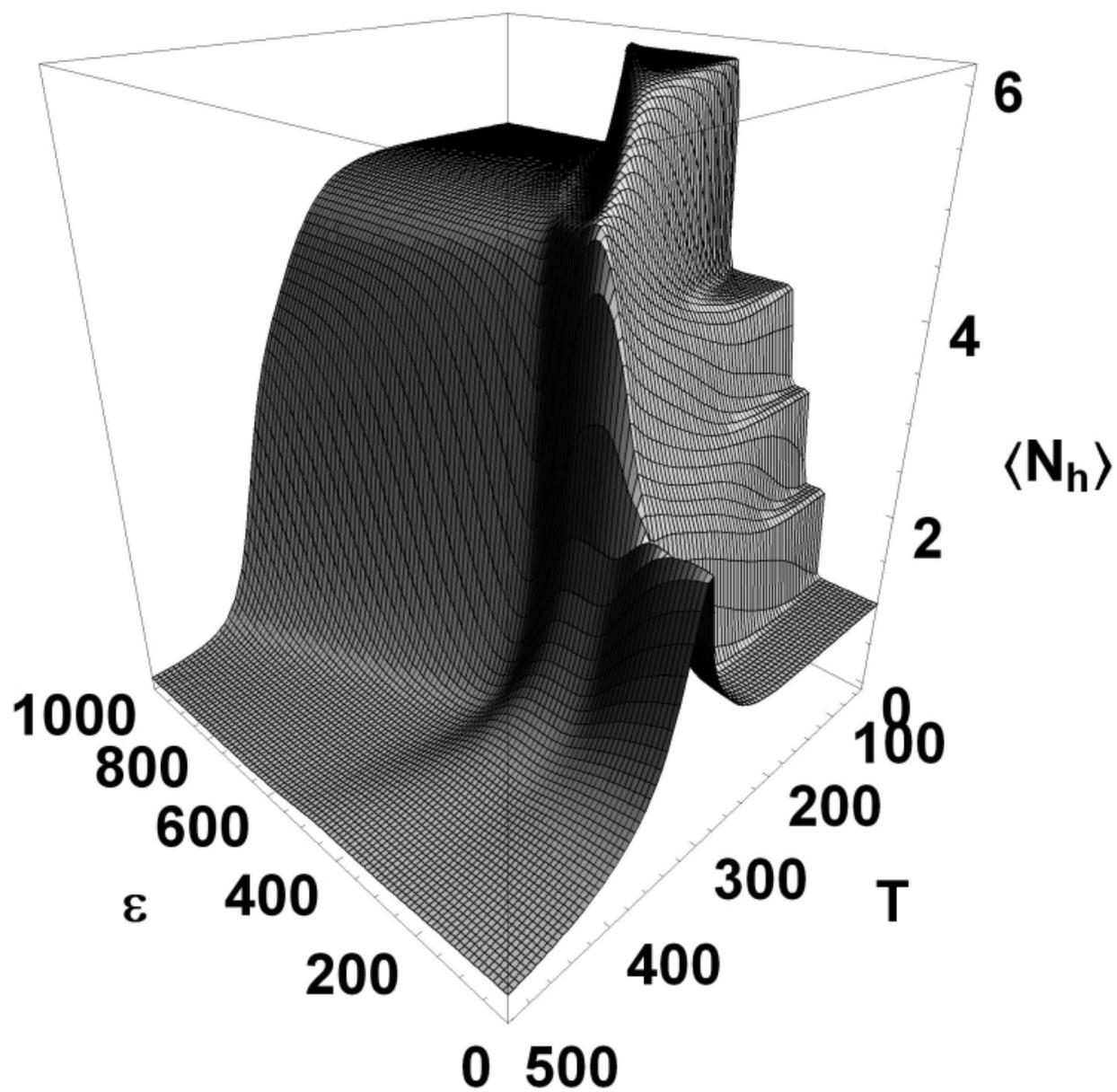

**Figure 4.** ⟨N_H⟩ as a function of ε and temperature for a chain with 60 beads.